# Functionalized Nanofullerenes for Hydrogen Storage: A Theoretical Perspective

N.S. Venkataramanan,[a,b] A. Suvitha,[b] H. Mizuseki,[b] Y. Kawazoe[b]

[a] College of Science,

California State University,

25800 Carlos Bee Boulevard,

Hayward, CA 94542.

Email: nsvenkataramanan@gmail.com, ramanan@imr.edu

[b]Institute for Materials Research (IMR)

Tohoku University,

2-1-1, Katahira, Aoba-ku, Sendai – 95035, Japan

**CONTENTS**

**Abstract :**

The increase in threats from global warming due to the consumption of fossil fuels requires our planet to adopt new strategies to harness the inexhaustible sources of energy. Hydrogen is an energy carrier which holds tremendous promise as a new renewable and clean energy option. Hydrogen is a convenient, safe, versatile fuel source that can be easily converted to a desired form of energy without releasing harmful emissions. However, no materials was found satisfy the desired goals and hence there is hunt for new materials that can store hydrogen reversibly at ambient conditions. In this chapter, we discuss and compare various nanofullerene materials proposed theoretically as storage medium for hydrogen. Doping of transition elements leads to clustering which reduces the gravimetric density of hydrogen, while doping of alkali and alkali-earth metals on the nanocage materials, such as carborides, boronitride, and boron cages, were stabilized by the charger transfer from the dopant to the nanocage. Further, the alkali or alkali-earth elements exist with a charge, which are found to be responsible for the higher uptake of hydrogen, through a dipole– dipole and change-induced dipole interaction. The binding energies of hydrogen on these systems were found to be in the range of 0.1 eV to 0.2 eV, which are ideal for the practical applications in a reversible system.

1. **Introduction**

Fossil fuels represent a vital energy resource for human activities and there are growing concerns that oil reserves cannot be sustained in the face of increasing worldwide demand. The increasing level of carbon dioxide ($CO_2$) in the atmosphere produced from the burning of fossil fuels has also raised increasing concerns over the impact on global eco-systems [1]. Ongoing research efforts to address these concerns have focused on many aspects including capture and storage of $CO_2$, and the use of cleaner energy sources such as methane ($CH_4$) or methanol as alternatives to petroleum or diesel in vehicular applications. The potential use of dihydrogen ($H_2$) as an energy carrier in principle reduces or even eliminates $CO_2$ emissions entirely at the point of use. To realized "The Hydrogen Economy" [2–7] the prime goal is that of inventing safe, efficient and effective stores for $H_2$ gas, and to replace current technologies based around the compression of $H_2$ as a liquid or as a gas using cryogenic temperatures or high pressures [7]. Therefore, there is major world-wide interest in meeting the United States, Department of Energy (DOE) targets of 6.5 wt% gravimetric and 45 g $L^{-1}$ volumetric $H_2$ storage by 2010, and 9.0 wt% and 81 g $L^{-1}$ by 2015 for mobile applications [8]. For all of these concepts to be realized in the applications the key target and required technological advance is the invention of new functional materials that are highly efficient in $H_2$ reversible storage.

The storage of gas in solids is currently a technology that is attracting great attention because of its many important applications [9–12]. Perhaps, the most well-known current area of research centers on the storage of hydrogen for energy applications, with viable energy storage for the hydrogen economy as the ultimate goal [13]. There are several reasons why one might want to store hydrogen inside a solid, rather than on a tank or cylinder. First, it is relatively common for more gas to be stored in a given volume of solid than one can store in a cylinder, leading to an increase in storage density of the gas. Second, there may be safety advantages associated with storage inside solids, especially if high pressures can then be avoided. Finally, small amounts of gases are actually easier to handle when stored in solids.

In the last 20 years, computer simulation studies on materials have contributed significantly to advance our understanding the mechanism and physics behind the materials, which is at the same time of fundamental scientific interest and of great technical importance [14]. This progress has been possible, on the one hand, because of improved simulation algorithms and the invention of powerful computers. Computational chemistry can be regarded as the application of chemical, mathematical, and computing skills to the solutions of chemical problems. Obtaining approximate solutions to the Schrödinger equation is the basis for most of the computational chemistry performed today. The quantum-chemical applications performed serve many times as source of inspiration for new methodological developments [15]. Among the quantum-chemical methods Density Functional Theory (DFT) is nowadays one of the most popular methods for ground state electronic structure calculation because of the favorable balance between accuracy and computational efficiency. Hence in recent years, DFT has been used to predict and realized new materials and to understand the properties of materials that can store hydrogen reversibility.

Physisorption of dihydrogen within a light, porous and robust material is an especially attractive option since this maximizes the possibility of highly reversible gas storage with fast kinetics and stability over multiple cycles. Although physisorption of $H_2$ in porous hosts can be highly reversible *via* changes in pressure and/or temperature, the storage involves low binding energies and isosteric heats of adsorption (typically less that 6 kJ mol$^{-1}$ ) and therefore cryogenic temperatures, typically 77 K, need to be used to achieve reasonable substrate uptake capacities.

In contrast, chemisorption of $H_2$ involves a much higher enthalpic Dyads contribution but may also lead to slower and poorer kinetics due to the requirement for reversible cleavage and formation of the H–H bond with concomitant generation of heat [16, 17]. Hence, there is an urgent need to find materials that has energy intermediate to physisorption and chemisorption. From the thermodynamic point of view, it has been proposed, those materials with hydrogen binding energy in the range of 0.2–0.5 eV would be ideal for the reversible $H_2$ storage [18].

The discovery of C60 (buckminsterfullerene) by Kroto et al. in 1985 has stimulated several researchers to search for similar structures for other inorganic compounds [19]. Hence in recent years the use of fullerene materials for hydrogen storage has gained attention. There exist several advantages in using fullerene for the hydrogen storage application [20]. The synthesis of nanotubes in pure form and with definite dimensions is difficult, whereas fullerenes can be prepared with high purity. Further, their curvature helps to avoid clustering thereby helping to keep the doped metals in isolated form.

Recently computational studies, involving both electronic-structure methods (*ab initio* and density functional theory) and classical molecular mechanics methods have added insight to these remarkable materials and the mechanisms of hydrogen adsorption [21]. The purpose of the review is to examine how effectively, theoretical methods have helped to elucidate the nature of interactions fullerene materials and the hydrogen molecules. In particular discussions were made on functionalization of fullerene like materials, and how the state of the art DFT methods have aided to the design of new novel nanomaterials.

## 2. Hydrogen storage in $C_{60}$

### 2.1. *Endohedral hydrogen storage in $C_{60}$ and carbon cages*

Initial experimental reports showing high levels of hydrogen storage were encouraging [22]. Hence, much attention has been devoted in several carbon compounds. Experimentally, crystalline *fcc* $C_{60}$ has been observed to absorb dihydrogen in octahedral interstices, providing a small storage capacity of only 0.28 wt% at 40°C [23]. Studies on $C_{60}$ molecules showed that the endohedral cage is electron density rich then the surface. To understand the adsorption nature and capacity of hydrogen on $C_{60}$ several theoretical studies have been carried out. To know the possibility of endohedral doping of hydrogen molecule in the $C_{60}$, Erkoç *et al*. carried out a semi-empirical molecular orbital calculation [24. 25]. Their calculations indicate that 24 hydrogen molecules can be placed inside the cage, and are highly endothermic. However later Dodziuk based on, MM+ calculations claimed that $C_{60}$ cannot accommodate more than one hydrogen molecule [26]. Latter, Narita and Oku concluded that $H_2$ is stable inside the $C_{60}$ cage at ambient conditions through a molecular dynamic study [27]. They also found that dihydrogen molecule could not escape from the cage due to the high activation barrier of 16 eV for penetration through the cage.

The disadvantage of using semi-emperical methods are they cannot provide correct insight in to the chemical bonding as they don't consider the electronic delocation of electron clouds in π-rich systems. Barajas-Barraza and Guirado-Lopez carried out a semi-empirical as well DFT calculations to analyze the hydrogen storage behavior in spheroidal $C_{60}$ and $C_{82}$ molecules. They observed that hydrogen molecules are stable inside the cage and chemisorption occurs on the outside wall of the cage. They propose that up to 19 $H_2$ molecules can be trapped inside a $C_{110}$ cage [28]. To understand the nature of interaction that exists between endohedral hydrogen molecule and the $C_{60}$, Liew *et al.* carried out calculation with the combination of PM3 semi-empirical and *ab initio* DFT method for geometrical optimization and for energy calculations respectively [29]. Their study showed that the hydrogen–hydrogen repulsive energies are weaker than the hydrogen and the all of the $C_{60}$ cage. It is further established that the repulsive energy among the $H_2$ molecules is not purely a function of the number of encapsulated hydrogen molecule. Recently, Farajian, *et al.* carried out a model study on the fullerene nanocage filled with hydrogen with a combined DFT and *ab initio* Molecular Dynamics (MD) [30]. They found that a maximum of 58 hydrogen atoms can be inside the $C_{60}$ cage in a metastable state and 20 of them are found to be chemisorbed on the inner surface of the cage. The hydrogen storage capacity of H58@C60 is approximately 7.5 wt %, which formally exceeds the U.S. Department of Energy (DOE) target. *Ab initio* MD studies at room temperature showed that cage opening occurs as shown in Figure 1. It was shown that the hydrogen chemisorption, which weakens the fullerene C-C bonds, plays the key role in the opening of the nanocage.

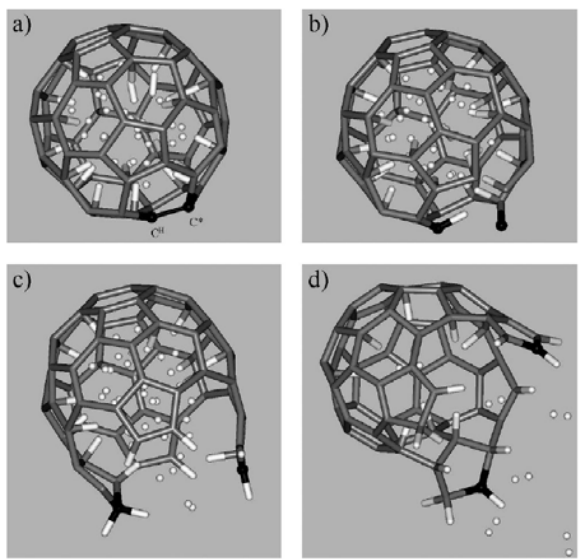

Figure 1. Snapshots of *ab initio* molecular dynamics simulations of H58@C60 structure at 300 K: (a) 0 fs, (b) 250 fs, (c) 350 fs, and (d) 800 fs. Breaking of bond between two carbon atoms shown in black initiates cage opening. (Reproduced with the permission of American Chemical Society from the reference 30)

## 2.2. *Transition decoration on $C_{60}$ for hydrogen storage*

Because of the weak binding energy and low storage capacity of carbon materials attention recent studies have focused to increase the binding strength of molecular hydrogen by adsorption of transition metals such as Ti onto the surfaces [31, 32]. Zhang *et al.* proposed transition metals (Sc, Ti, V, Cr, Mn, Fe, Co and Ni) as the adsorbents on $C_{60}$ as a possible hydrogen storage material [33]. They found that Sc doped on cyclopentadiene ring was able to hold up to 5 hydrogen molecules in dihydride from. The calculated binding energy is about 0.3 eV/$H_2$ with a theoretical maximum storage density of ~ 9 wt % on the $C_{60}$ system. Latter, Yildirim *et al.*. made a theoretical study by decorating transition elements (Sc, Ti, V, Cr, Fe, Co and Mn) on the $C_{60}$ fullerene [34]. Heavier metals such as Mn, Fe and Co do not bond on $C_{60}$, while Ti was found have a good adsorption energy and binding energy towards hydrogen molecules. Each doped Ti atom was found to bind up to 4 hydrogen molecules with an average binding energy of about 0.465 eV per dihydrogen molecule. They proposed that up to 14 Ti atoms can be decorated on the $C_{60}$ which has approximately 7.5 wt % storage capacity. Latter Kang *et al.* proposed Ni decoration on the $C_{60}$ cage [35]. Their calculations shows that Ni prefers the edge site between two hexagonal rings and each Ni atom was could adsorb 3 $H_2$ molecules. They decorated the $C_{60}$ with 30 Ni atoms, which could store 90 hydrogen molecules as shown in Figure. 2. Thus they propose a hydrogen storage capacity of 6.8 wt % on the Ni doped $C_{60}$ cage.

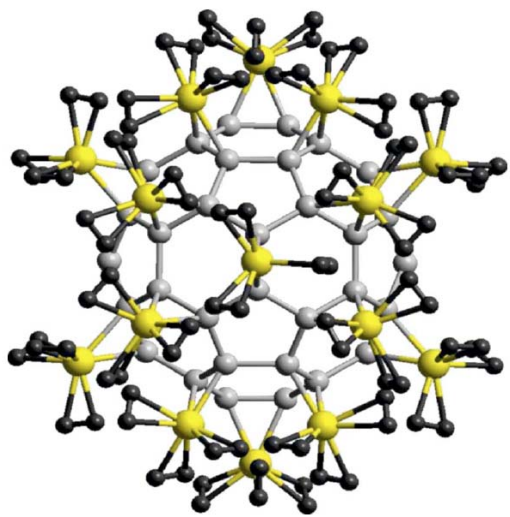

Figure 2. Proposed structure with 30 Ni atoms decorated on the $C_{60}$ with 3 $H_2$ adsorbed on each Ni atom. (Reproduced with the permission of American Physical Society from reference 35)

## 2.3. Alkali and Alkali earth atom doping on $C_{60}$ for hydrogen storage

Recently, theoretical studies showed that transition elements decorated on to the surfaces of nanostructures undergo clustering during the reversible hydrogenation [36-38]. Hence, attention has been focused on the alkali atom doping. A lot of experimental work was performed to investigate the hydrogen adsorption in alkali doped carbon nanotubes and MOF's with the aim of improving the storage capacity of these porous materials [39, 40]. The reason for choosing alkali atoms for decoration is of three-fold reasons. First is the alkali atoms are light weight elements compared to the transition elements helping thereby providing a way to achieve high gravimetric density. Secondly, $C_{60}$ fullerene has a large electron affinity and the doped alkali element would partially donate its valance electron resulting in a strong bonding. Thirdly, after donation the alkali atoms remain in cationic state, which can binding hydrogen in molecular from due to the polarization mechanism [41, 42]. In the first case, Sun *et al.* using the Vienna *ab initio* simulation (VASP) package performed an in-detail study on the Li doping on the $C_{60}$ molecule and their hydrogen adsorption using GGA method and a pseudopotential plane wave basis set [43]. They decorated 12 Li atoms on the $C_{60}$ by placing Li atoms on the pentagonal rings as shown in Figure 3(a). The calculated adsorption energy of Li atom to $C_{60}$ is 1.8 eV which is larger than the $Li_2$ dimer energy and the Li bulky cohesive energy.

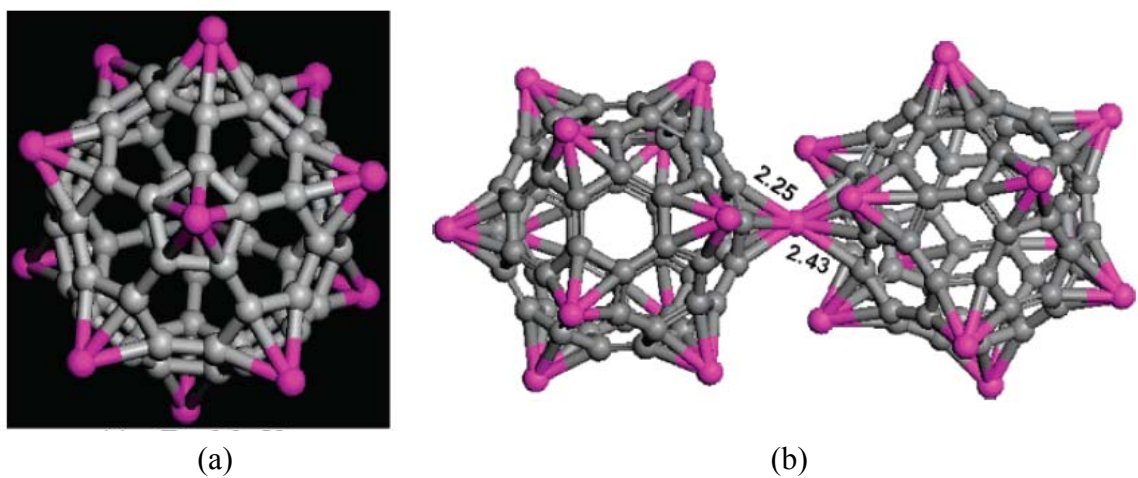

(a)            (b)

Figure 3. Optimized structure of (a) 12 Li atoms on $C_{60}$ (b) $Li_{12}C_{60}$ dimer (Reproduced with the permission of American Chemical Society from reference 43)

To know the possibility of clustering, between the two intermolecular Li doped $C_{60}$ fullerene, calculations were carried out. As the Li atoms exist in charged state, they are repelled away from one another and are kept at a distance of 3.34 Å. They also examined the Li atom linking two $Li_{12}C_{60}$ cluster as shown in Figure. 3(b) and its hydrogen adsorption properties. They found such a possibility would reduce the hydrogen storing capacity from 60 to 40 hydrogen molecules and thereby yielding an approximate gravimetric density of hydrogen of 9 wt %.

In 2008, Chandrakumar *et al.* using the hybrid B3LYP functional together with the 6-31G(d,p) basis set as implemented in GAMESS program and studied the interaction of alkali atoms (Li, Na and K) on $C_{60}$ and their hydrogen adsorption properties [44]. They observed the alkali atoms prefer to adsorb on the hexagonal ring and up to eight alkali atoms were found to adsorb on the $C_{60}$ cage, which was in sequence with an earlier experimental prediction [45]. They observed that the adsorption energy of the alkali metal atoms and charge transfer are not affected by the number of alkali atoms. Hence in their calculations they considered the system to be charged. Hydrogen adsorption of alkali doped $C_{60}$ shows that each Li can hold up to two $H_2$ molecules, whereas sodium and potassium can hold up to 8 $H_2$ molecules. Thus they propose, the sodium doped $C_{60}$ can have a gravimetric density of ~ 9.5 wt%. The binding energy calculation shows sodium has a better binding energy of 2.04 kcal mol$^{-1}$ than potassium cation.

In 2008, Yong *et al.* in a communication, propose calcium as the substitute for the alkali elements as they provide a weak binding energy for hydrogen molecules [46]. They attempted doping with Ca, Sr, Be and Mg, however, only Ca and Sr was found to adsorb strongly on the hexagonal and pentagonal rings of $C_{60}$ surface. The strong binding for Ca and Sr is attributed to an intriguing charge transfer mechanism involving the empty *d* levels of the metals. They found that as many as 32 Ca atoms can be coated on the $C_{60}$ surface and the charge redistribution that occurs during the Ca binding gives rise to electric fields surrounding the collated fullerenes, which can help to adsorb the hydrogen molecule. Each Ca atom was found to hold up to 5 hydrogen molecules in near molecular form. Further, the calculated binding energy by LDA methods is higher than GGA method. They propose a hydrogen uptake of 8.4 wt% with a binding energy of ~ 0.4 eV per dihydrogen molecule on these materials.

Latter, Sun *et al.* carried out a MD study to know the stability of the Ca coated $C_{60}$ [47]. They found that Ca coated $C_{60}$ material is highly stable. Hydrogen adsorption on these material shows that first 30 $H_2$ molecules dissociate and bind automatically on the 60 triangular faces of the Ca coated fullerene with an average binding energy of 0.45 eV per hydrogen atom. In addition a second layer of 32 Hydrogen molecules were found to bind in quasi molecular form with an average binding energy of 0.11eV per dihydrogen molecule. The optimized geometry with full hydrogen storage capacity is shown in Figure 4. The proposed gravimetric density of this material is 6.2 wt%.

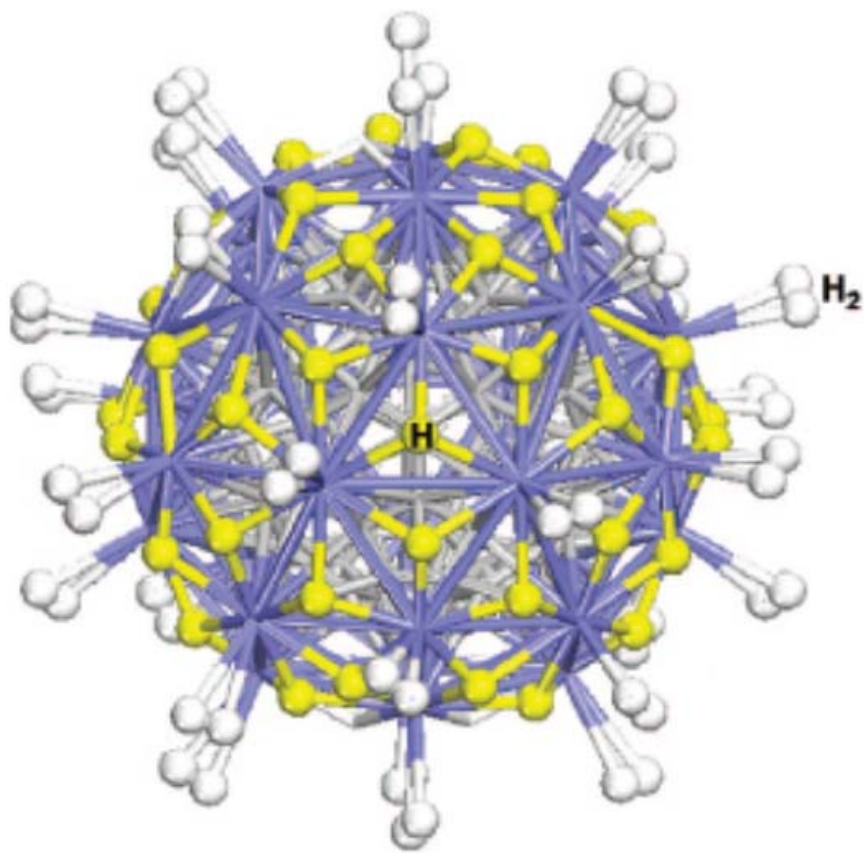

Figure 4. Fully optimized geometry hydrogen saturated on Ca coated $C_{60}$. The yellow balls represent the hydrogen in hydride from. (Reproduced with the permission from American Chemical Society from reference 47)

### 3. Hydrogen storage in BN fullerenes

#### *3.1. Endohedral storage*

Since the hydrogen adsorption on carbon materials have been controversial, attention has been focused on non-carbon materials for hydrogen storage [48. 49]. BN materials are considered to be analogous to carbon materials, but have higher stability. Moreover, they have hetropolar nature, due to which they binding hydrogen more strongly than carbon materials. Further, it has been recently established hydrogen storage in the form of chemical hydrides can be reversibly stored on these materials in the presence of suitable catalysts. Wang *et al.* have showed that *h*-BN materials prepared by mechanical milling under hydrogen atmosphere could reach 2.6 wt % after milling for 80 h [50]. Latter, Tang *et al.* reached 4.6 wt % of hydrogen on the collapsed BN nanotubes [51]. Oku *et al.* found that BN fullerene are better candidates for hydrogen storage reaching storage capacity of ~3.2 wt % [52, 53]. Further, their theoretical calculations showed a minimum energy barrier of 14 eV to pass though hexagonal rings to the endohedral site. Their molecular orbital calculations endohedral hydrogen molecules are more stable in BN clusters than $C_{60}$ and can store up to 4.9 wt% in molecular form [54].

Recently Sun *et al.* carried out an *ab intio* DFT and MD studies to understand the stability of hydrogen molecules inside the $B_{36}N_{36}$ cage structure [55]. They found that hydrogen molecules enter though the hexagonal face of the $B_{36}N_{36}$ cage and prefers to remain inside the cage in the molecular form. The calculated energy barrier for the hydrogen molecule to enter and to escape from the cage is 1.40 and 1.52 eV respectively. At zero temperature up to 18 hydrogen molecules can be stored inside the can which corresponds to 4 wt %. When more hydrogen molecules are introduced the cage structure was found to be collapsed. Further their MD studies shows hydrogen molecules escapes out of the cavity at 300 K, thereby making these materials less likely to be used for practical applications.

The hydrogen adsorption studies by endohedral metal atom (Li, Na, Be , Mg and Ti) doped $B_{36}N_{36}$ cage was recently studied by Deng and co-workers [56]. Doping Li, Ti and Be were found to be exothermic and are feasible, whereas Na and Mg are found to form a metastable complex. Further, after doping Ti atom was found to carry a high magnetic moment of 4.0 $\mu_B$ without a spin transfer to the cage [57]. Hydrogen adsorption on the $B_{36}N_{36}$ cage shows that the formation of N-H bond is endothermic and the B sites are electron-deficient, implying that B sites are more favorable than the N sites for H chemisorption. Adsorption studies shows that it is energetically favorable for five hydrogen molecules to enter the Ti doped $B_{36}N_{36}$ caged and form a Kubas complex with the endohedral Ti atom. Thus the total number of hydrogen that can be attached to the Ti doped $B_{36}N_{36}$ cage enhancing the total hydrogen storage in $B_{36}N_{36}$ up to 8 wt %.

### 3.2. *Alkali doped BN cage for hydrogen storage*

As the chemical hydrides needs high activation energy during their release, attempts were made to store hydrogen in molecular from. Recently, we have made in detail study to understand the hydrogen adsorption of alkali atom doped BN fullerene in the molecular form using the pseudo potential as well as with Gaussian type orbital basis set [58]. Out of the possible eight sites of doping, the bridge site near the tetragonal ring was found to have the highest binding energy. Bader changer transfer analysis indicates that about 0.82 *e*, 067*e* and 0.58*e* resides on the Li, Na and K atoms respectively after doping at the bridge sites. Figure 5 (a) and (b) shows the contour plot for the excess and depletion charge after doping the Li atom on the BN cage. It is evident from the figure that the charge on the Li atom is depleted and is in excess near the BN bond. The significant positive charge on alkali-metal atoms allows the possibility that hydrogen atoms may be bound to the alkali-metal center though electrostatic-charged quadrupole and charge-induced dipole interactions in quasi-molecular form. The calculated binding energy and the average Bader and Natural bonding analysis changes for the alkali atom doped $B_{36}N_{36}$ fullerene cage is shown in Table 1.

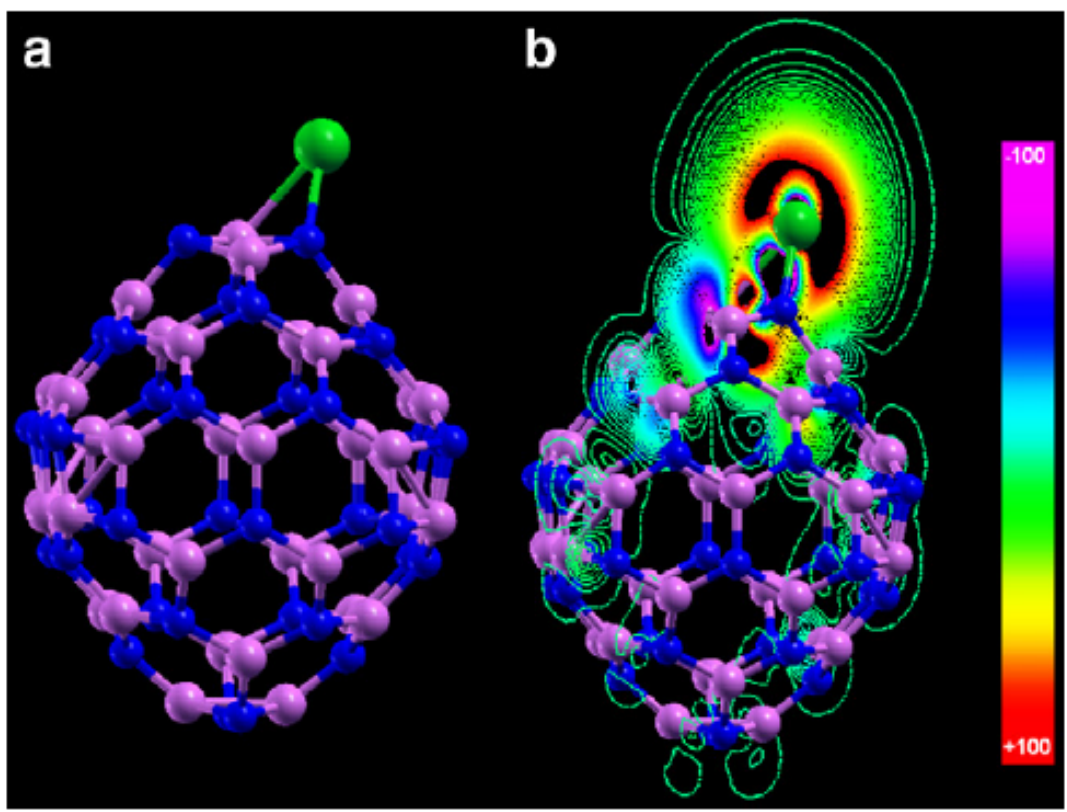

Figure 5. (a) Geometry optimized structure for the Li doped on $B_{36}N_{36}$ fullerene. (b) Contour plot of the excess and depletion charge (e/Å$^3$) for the Li doped $B_{36}N_{36}$ fullerene cage in the plane containing the Li atom and the bridged BN bond.

Table 1. Geometrical parameters (Å) adsorption energy of alkali atoms (eV) and Bader charge on the alkali-metal atom doped on $B_{36}N_{36}$ cage in the most stable configuration. [a] calculated with pseudo potential. [b] with Gaussian type orbital basis set.

| System | Mean bond length (Å) | | | | BE (eV) | Average Bader charge on $X^a$ | NBO charge on $X^b$ |
| --- | --- | --- | --- | --- | --- | --- | --- |
| | $B-X^a$ | $N-X^a$ | $B-X^b$ | $N-X^b$ | | | |
| $LiB_{36}N_{36}$ | 2.137 | 2.178 | 2.211 | 1.907 | -0.725 | 0.82 | 0.85 |
| $NaB_{36}N_{36}$ | 2.754 | 2.626 | 2.455 | 2.635 | -0.283 | 0.67 | 0.65 |
| $KB_{36}N_{36}$ | 2.869 | 3.044 | 2.766 | 2.960 | -0.369 | 0.58 | 0.41 |
| $Li_6B_{36}N_{36}$ | 2.250 | 2.200 | - | - | -2.86 | 0.74 | - |
| $Na_6B_{36}N_{36}$ | 3.040 | 2.847 | - | - | -1.50 | 0.52 | - |
| $K_6B_{36}N_{36}$ | 3.021 | 3.246 | - | - | -1.40 | 037 | - |

The calculated average binding energy for the six Li atom doped $B_{36}N_{36}$ fullerene cage was higher than their clustering energy and the bulk Li cohesive energy of 1.7 eV. This indicates that the doped metal atoms will stay intact on the bridge sites. We than focused on the adsorption of hydrogen molecules by the alkali metal cation doped $B_{36}N_{36}$ cage. Upon doping the first hydrogen atom near the lithium, we found that it is bound molecularly with a binding energy per $H_2$ of -0.219 eV and with a slightly stretched H–H bond distance of 0.757 Å. The mean distance between the Li and $H_2$ is 2.07 Å. When we increase the number of hydrogen molecules to two, the binding energy decreases to -0.210 eV and the $H_2$ bond length becomes 0.756 Å. To know the maximum number of hydrogen molecules a lithium atom can bind, we doped three to four hydrogen molecules near the lithium atom and optimized the geometry without symmetry restrictions. We found that three $H_2$ molecules are bound to the lithium atom and the fourth $H_2$ is moved to a distance of 3.881 Å from the Li atom. Thus, each Li atom was found to hold up to three $H_2$ molecules in quasi-molecular form, and the calculated binding energy decreased to -0.189 eV, along with a decrease in H–H bond length to 0.755 Å. The calculated binding energy for the Li and other alkali doped systems are shown in Table 2. If the $Li_6B_{36}N_{36}$ can adsorb 18 dihydrogen molecules the total gravimetric density of molecular hydrogen in $Li_6B_{36}N_{36}$ can reach 3.7 wt.%. It is important to point out here that the pure generalized gradient approximation (GGA) method underestimates the binding energy [59]. Hence in our system the binding energy may be higher than the calculated results

Table 2. Geometrical parameters (Å), binding energy per $H_2$ ($\Delta En$) and for the three hydrogen molecules adsorbed on each alkali-metal atom doped $B_{36}N_{36}$ cage.

| System | Mean bond length (Å) | | | H-H bond length (Å) | $\Delta En$ (eV) |
|---|---|---|---|---|---|
| | B-X | N-X | Li-$H_2$ | | |
| $LiB_{36}N_{36}(H_2)_3$ | 2.217 | 2.287 | 2.150 | 0.755 | -0.189 |
| $NaB_{36}N_{36}(H_2)_3$ | 2.826 | 2.796 | 2.470 | 0.755 | -0.175 |
| $KB_{36}N_{36}(H_2)_3$ | 2.997 | 3.214 | 2.968 | 0.752 | -0.021 |
| $Li_6B_{36}N_{36}(H_2)_3$ | 2.221 | 2.316 | 3.021 | 0.751 | -0.146 |

To understand the nature of interaction between the Li atom and hydrogen molecules we calculated the Bader charge on the hydrogen molecules. We found an average charger of 0.96 *e* on the hydrogen atoms, indicating a transfer of charge from the hydrogen molecule to the Li atom which stabilizes the complex formation. To further confirm the charge transfer we plotted the excess and depletion charge for the hydrogen adsorbed system. From Figure 6 (a–c), it is clear that the hydrogen atoms are polarized by the positively charge Li atom. It is worth pointing out here, that $H_2$ has zero dipole moment, but has a large quadrupole moment and high polarizability. Thus, these studies clearly indicate that the interaction between hydrogen and the alkali center to be dipole – quadrupole and dipole – induced dipole electrostatic in nature.

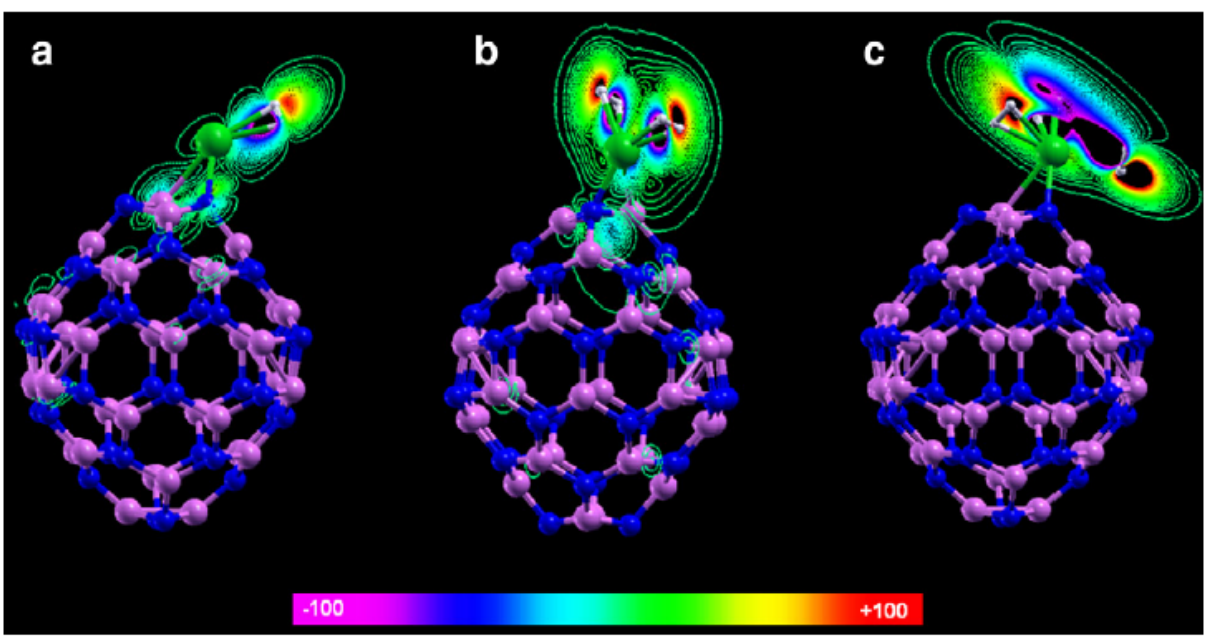

Figure 6. Contour plot for the excess and depletion charge (e/Å3) for the Li doped $B_{36}N_{36}$ fullerene cage with (a) one hydrogen molecule, (b) two hydrogen molecules, (c) three hydrogen molecules in the best plane of view.

To know the feasibility of chemisorption, we placed hydrogen atoms on the undoped 30 B and 30 N sites and fully optimized the structure. With the addition of hydrogen on the B–N cage, a slight structural deformation was observed with an elongation in the BN bond lengths from 1.48 to 1.64 Å and a concomitant change in bond angles. These are attributed to the hybridization change from *sp*2 to *sp*3 state. The calculated chemisorption energy for the hydrogen chemisorption on the N site was +0.249 eV while on the B site was -0.407 eV. The average chemisorption energy for the hydrogen atom on the thirty undoped B and thirty undoped N sites was -2.13 eV. This is close to the previous prediction for the corresponding chemisorption energy of the pristine $B_{36}N_{36}$ cage [59]. The negative value of the chemisorption energy implies that the chemisorption is an exothermic process. Further, the average Bader charge on the Li atom was 0.61*e* indicating that still they can attract hydrogen molecules. To know the possibility of hydrogen adsorption by the hydrogenated Li doped BN fullerene we placed, three hydrogen molecules on each of the Li atom and optimized the geometry. After full relaxation, we found that two hydrogen atoms are bound to the Li atom, while the third hydrogen molecule is moved away from the Li atom. The calculated binding energy per $H_2$ for the molecular hydrogen was -0.147 eV. Thus the maximum hydrogen storage capacity of Li doped BN fullerene is 8.9 wt.% in which sixty(60) hydrogen atom were chemisorbed and twelve(12) $H_2$ were adsorbed in molecular form and the binding energies are far lower than the Li C60 decorated fullerene system.

## 4. Hydrogenated silicon fullerene as hydrogen storage material

Studies of hydrogen interaction with silicon fullerene have led to the prediction of new novel materials with empty cage $Si_nH_n$ fullerenes in which hydrogen atoms are used to terminate the dangling bonds of each Si atom in the fullerene cage [60. 61]. This work stimulated researchers to know the possibility of Si fullerenes as hydrogen storage materials. In the year 2007, Ma *et al.* made a DFT study on the hydrogen storage possibility on the $Si_{60}$ fullerene using Dmol$^3$ and Gaussian G03 codes [62]. They found that $Si_{60}$ after adsorbing 60 hydrides on the surfaces stable and the hydrogenated silicon fullerene maintains the $I_h$ symmetry (Figure. 7). They have demonstrated that up to 58 hydrogen molecules can be stored into the $Si_{60}H_{60}$ cavity and the use of pure functional PW91 underestimate the binding energy than hybrid B3LYP method. They propose a total gravimetric density of hydrogen both including the chemisorbed and the endohedral hydrogen molecules to 9.48 wt %.

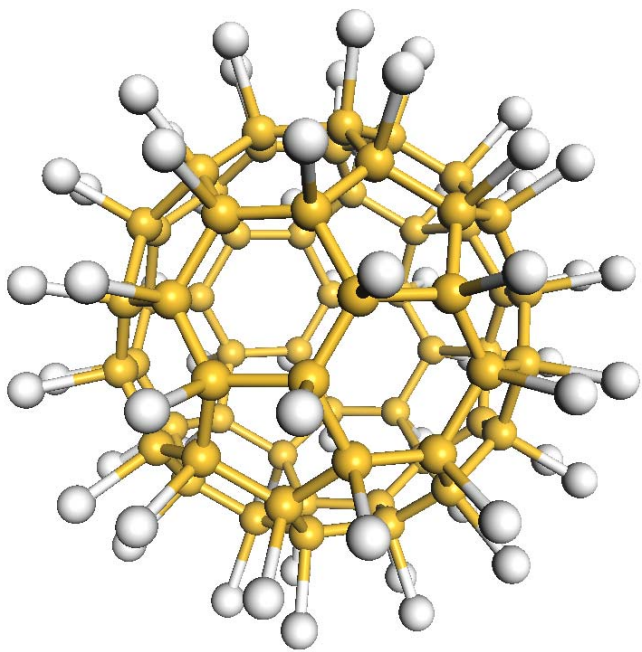

Figure 7. Perspective view of $Si_{60}H_{60}$ structure. Yellow ball represents Si atom and white balls represents hydrogen molecules.

Recently, Das *et al.* made a prediction on the hydrogen storage possibility on the $Si_{60}H_{60}$ decorated with titanium atoms [63]. Titanium atoms were able to be capped exohedrally on the hexagonal faces in cationic form, enabling storage of hydrogen in the molecular state. Up to 4 hydrogen molecules in the near molecular form were able to adsorb on the Ti atom doped with adsorption energy decreasing from 0.47 eV to 0.26 eV for the first and fourth hydrogen molecule respectively. Since, the transition elements such as Ti suffers due to the clustering ability, they propose phosphorus doping similar to the boron doping in the $C_{60}$ fullerene [64]. By doping P atoms they found that clustering tendency of Ti atoms can be avoided and propose a new materials $P_{10}Si_{50}H_{50}$ as shown in Figure 8 for hydrogen storage. The decoration of $P_{10}Si_{50}H_{50}$ with Ti atoms was found to be capable to store up to 5.23 wt % of hydrogen.

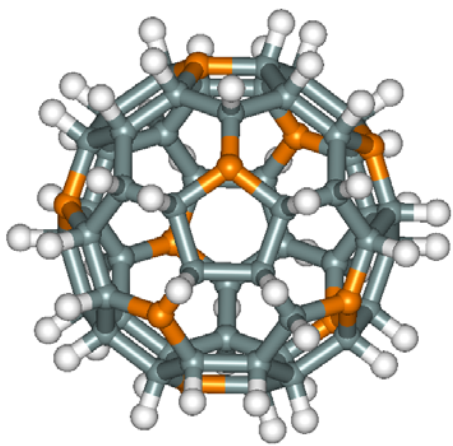

Figure 8. Optimized structure of $P_{10}Si_{50}H_{50}$ fullerene. Gray ball represent Si atom, Organic balls represents P atom and White represents H atom.

**5.     Alkali, alkali –earth and transition Metal, doped boron cage for hydrogen storage**

Recently, Yakobson et al predicted the existence of an unusually stable new boron cage with 80 boron atoms as shown in Figure 9 [65, 66] The shape of the cluster is very similar to that of the well-known $C_{60}$ fullerene, but has an additional atom in the center of each hexagon. This made several researchers to consider boron bucky ball as the hydrogen storage media. In 2007, Zhou *et al.* carried out a feasibility study of the alkali-metal (Li, Na and K) doped $B_{80}$ fullerenes for hydrogen storage within the framework of DFT method using the Dmol$^3$ package [67]. Out of the possible six sites of doping they found that alkali elements prefer the center of the pentagon ring. Further the positive charge of the alkali atoms increases in the order of Li, Na and K. Hydrogen adsorption studies on the alkali doped $B_{80}$ cage shows that Li atom can hold only 2 hydrogen molecule, whereas the Na and K atoms are capable to adsorb six and eight hydrogen molecules respectively. They have identified 12 pentagon sites for the alkali elements doping and $B_{80}Na_{12}$ and $B_{80}K_{12}$ system can store 72 $H_2$ molecules, correspond to the gravimetric density of 11.2 and 9.8 wt % respectively.

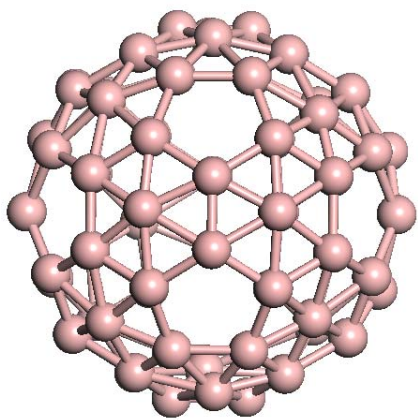

Figure 9. $B_{80}$ Boron bucky ball

Recently, Wang *et al.* have investigated the hydrogen adsorption capacities of light and transition metal (Li, Na, K, Be, Mg, Ca, Sc, Ti and V) coated on $B_{80}$ cage [68] The calculated binding energy of hydrogen on these doped systems shows that alkali elements have low binding energy. Further, only the Ca, Sc and Ti elements are capable to bind six $H_2$. Calculation carried out with Ca and Sc cluster adsorbed on $B_{80}$ shows that they have high energy and hence they propose the Ca and Sc coated materials are highly stable. They also notice a high dipole moment on the Ca doped $B_{80}$, which polarizes the $H_2$ molecules and dipole-dipole as well as the charge-dipole interaction, are responsible for the binding of $H_2$ around the $Ca_{12}B_{80}$ cage, whereas in the case of Sc doped system, the Dewar-Kubas interaction was the factor that binding $H_2$ molecules [69]. In the $Ca_{12}B_{80}$ with 12 Ca atoms coating on the pentagonal rings of $B_{80}$ can adsorb 66 $H_2$ molecules while, $Sc_{12}B_{80}$ system can hold up to 48 $H_2$ molecules which account for a storage capacity of 9.0 wt % and 7.9 wt % respectively. While a latter study by Chen *et al.* propose only 5 $H_2$ can bind to the Ca doped on the $B_{80}$ cage and the fully coated system with 12 Ca can store 60 $H_2$ molecules to a gravimetric density of hydrogen storage of 8.2 wt % [70].

## 6. Hydrogen storage on Boroncarbide cages

Recent studies have shown that boron substitution on carbon materials imparts stability to them and MD studies have indicated that they are stable up to 1000 K. It was reported experimentally that up to six boron atoms could replace carbon to form $C_{54}B_6$ cage [71]. In 2006, Zhang *et al.* studied the adsorption of $H_2$ molecules on the $C_{54}X_6$ (X = B, Be) cage [72]. By LDA method binding energy for $H_2$ was 0.39 eV, whereas, by GGA the energy was -0.03 eV, and the negative value implies repulsion between B and $H_2$ molecule. LDOS calculations show that the coupling between a localized empty $p_z$ orbital on the B and the $H_2$ σ orbital was helpful for the enhanced interaction. Further, in the $C_{54}B_6$ system the $H_2$ interaction energy increase up to first two hydrogen and then decreases and reaches the minima at six. Later, Ihm *et al.* have established that hydrogen adsorption this system undergo a trapping-mediated dissociative chemisorption on these systems [73]. They employed Polanyi-Wigner equation with the van't Hoff-Arrhenius law along with the MD simulation and found that $H_2$ dissociate on the B doped fullerene system even in the absence of any catalyst.

In 2008, Zhang *et al.* proposed that the organometallic nanostructure, $X_{12}B_{24}C_{36}$ (X= Ti, Sc) for the hydrogen storage [74]. They found on Ti doped system, first hydrogen adsorption takes place in dissociative form with a binding energy of 1.85 eV and five other $H_2$ in molecular form. On contrary, Sc doped system adsorbs six $H_2$ in molecular form with a binding energy of 0.40 eV/$H_2$ which is nearly ideal for practical purpose. The metal-carboride buckyball $Sc_{12}B_{24}C_{36}$ with and without 72 $H_2$ are shown in Figure 10. Further, the metal- metal distance in these systems are away from each other by 5.5 Å thereby reducing the possibility of clustering. They propose a gravimetric density of hydrogen storage of 10.5 wt % on this organometallic buckyball.

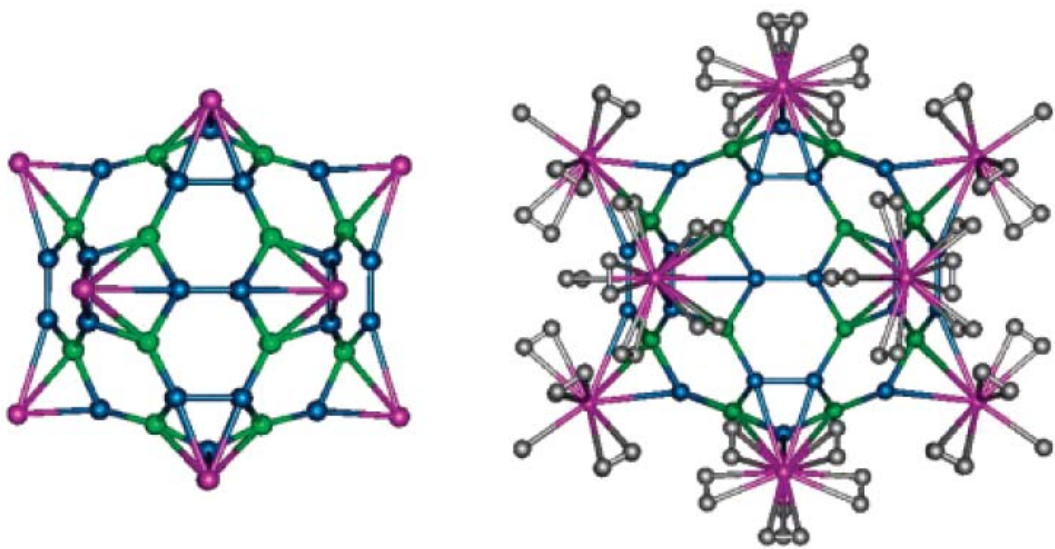

Figure 10. Structure of (a) $Sc_{12}B_{24}C_{36}$ and (b) $Sc_{12}B_{24}C_{36}$ with 72 $H_2$ atoms adsorbed on the organometallic fullerene.

In 2009, Sun *et al.* showed that 12 atoms of Li can be decorated on $C_{48}B_{12}$ system, where in Li atoms remain isolated and each of them are found to bind three $H_2$ molecules with a binding energies between 0.172 and 0.135 eV/$H_2$ [75]. To determine the geometry of $Li_{12}C_{48}B_{12}$ system they consider, four isomers and their calculated energies shows that Li in isolated form is more stable than other and Li atom prefers the hexagonal center in the system. Boron doping results in electron-deficient heterofullerene and this makes Li atoms to donate their 2*s* electrons, thereby remaining in a positively charged state which helps in binding hydrogen to the Li centers. The gravimetric density reaches 9 wt% and average binding energy per $H_2$ molecule is 0.135 eV in this system.

## 7. Hydrogen storage on metal nitrides

Interaction of metal ions with small molecules showed that free ions binds hydrogen molecules strongly, whereas ligands adsorbed on metal ions have moderate binding energy [76].

Hence recent study has been focused on metal nitride systems [77]. Especially Al-N cages are studies as Al and N are light in mass and Al atoms are dispersed in cages having no problem of metal clustering as encountered in metal decorated fullerenes [78]. The interactions of hydrogen molecules with $(AlN)_x$ cages (x = 12, 24, 36) was studied recently using GGA method with spin polarization as implemented in VASP program, based on the ultrasoft pseudopotentials and a plane-wave basis set. Hydrogen adsorption on the Al site shows that it can adsorb one hydrogen and the binding energy decreases with the increase in the size of the AlN cluster. When 12 $H_2$ molecules were introduced to a $(AlN)_{12}$ cage, one $H_2$ molecule for each Al site, the average binding energy was found to be 0.176 eV/$H_2$, while the Al–$H_2$ and H–H average bond lengths are slightly changed to 2.274 and 0.760 Å and the corresponding weight percentage is 4.7%. The average binding energy of H2 on $(AlN)_{12}$ system was 0.176 eV, whereas on the $(AlN)_{36}$ the energy decreases to 0.133 eV, while the orbital interaction and polarization are responsible for the interaction of H2 molecules with the cages.

## 8. Conclusion

Three main strategies are used to store the hydrogen on the fullerene like cages. Endohedrally doping was possible to a little extent in $C_{60}$ fullerene as its cage center has a higher electron cloud than on the surface. However, high pressure results in the chemical hydrogenation and cage opening. The second attempt the use of transition metal dopant's results in clustering and decreases the gravimetric density of materials. Thirdly, the use of light metal elements those including alkali and alkali-earth metals help to increase the hydrogen gravimetric density and increase in stability of the materials. The change induced dipole-dipole interactions are found to be responsible for the hydrogen interactions and higher uptake on these doped systems.

This theoretical modeling of the hydrogen storage provides a valuable understanding to experimentalist to develop new materials. However, the present theoretical methods were not efficient to predict the correct binding energy on the systems. Thus the future outlook for the theoretical chemists/physics will be focused toward the development of new methodologies, which provide correct binding energies for hydrogen.

**Acknowlegements**

This work has been supported by New Energy and Industrial Technology Development Organization (NEDO) under "Advanced Fundamental Research Project on Hydrogen Storage Materials". The authors thank the crew of the Center for Computational Materials Science at Institute for Materials Research, Tohoku University, for their continuous support of the HITACHI SR11000 supercomputing facility. NSV thanks all the collaborators who's name are provided in the reference section.